\documentclass[3p,times,procedia,option,numerical,natbib,linenumbers,superscriptaddress,nofootinbib,showpacs,showkeys,longbibliography]{elsarticle}

\usepackage{txfonts} 
\usepackage{nupha_ecrc}
\usepackage{amssymb}
\usepackage{hyperref}
\usepackage{amsmath,amssymb,amsfonts}
\usepackage{graphicx,color}
\usepackage[figuresright]{rotating}
\usepackage{lineno}
\usepackage{natbib}
\setcitestyle{square, comma, numbers,sort&compress, super}

\def \be {\begin{equation}}
\def \ee {\end{equation}}
\def \ee  {\end{equation}}
\def \bea {\begin{eqnarray}}
\def \eea {\end{eqnarray}}

\newcommand{\roots} {\mbox{$\mathrm{\sqrt{s_{NN}}}$}}

\newcommand{\pT} {\mathrm{p_{T}}}

\def \GeVc {\mbox{$\mathrm{GeV} / c $}}

\volume{00}

\firstpage{1}

\journalname{Nuclear Physics A}

\runauth{}

\jid{nupha}

\jnltitlelogo{Nuclear Physics A}

\usepackage{amssymb}

\usepackage[figuresright]{rotating}

\begin{document}

\begin{frontmatter}



\dochead{XXVIIIth International Conference on Ultrarelativistic Nucleus-Nucleus Collisions\\ (Quark Matter 2019)}

\title{Beam-energy and collision-system dependence of the linear and mode-coupled flow harmonics from STAR}


\author{Niseem Magdy for the STAR Collaboration\fnref{xxx}}
\fntext[xxx]{niseemm@gmail.com}
\address{Department of Physics, University of Illinois at Chicago, Chicago, Illinois 60607, USA}

\begin{abstract}
Recent measurements and hydrodynamic model calculations suggest that the higher-order flow coefficients $v_{4}$ and $v_{5}$ have two contributions: a linear contribution driven by the initial-state eccentricities, $\varepsilon_{n}$, and a mode-coupled contribution derived from the lower-order eccentricity coefficients $\varepsilon_{2}$ and $\varepsilon_{3}$. Measurements of these two contributions to $v_{4}$ and $v_{5}$ provide crucial insights to discern initial-state models and to constrain the temperature-dependent specific shear viscosity, $\eta/s$, of the plasma produced in heavy-ion collisions. In this work, we have employed the two-subevents cumulant technique to provide the first beam-energy and collision-system dependence of the linear and mode-coupled contributions to the higher-order flow harmonics.
 Our results are shown and discussed for several centrality intervals for U+U collisions at $\roots$= 193 GeV, Au+Au collisions at $\roots$=200, and 54.4 GeV and Cu+Au collisions at $\roots$=200~GeV. The results are compared with similar studies performed by the ALICE experiment at LHC.

\end{abstract}


\end{frontmatter}


\section{Introduction}
\label{Introduction}

Ongoing investigations of the matter produced in heavy-ion collisions at the Relativistic Heavy Ion Collider (RHIC) and the Large Hadron Collider (LHC) indicate that an exotic state of matter called Quark-Gluon Plasma (QGP) is produced.
Many of these studies are aimed to understand the dynamical evolution and transport properties of QGP \cite{Heinz:2001xi,Hirano:2005xf}.

 The measurements of the azimuthal anisotropy of the particle production called anisotropic flow have been used in various studies to explain the viscous hydrodynamic response to the initial spatial distribution in energy density,  created in the early stages of the collision \cite{Magdy:2017ohf,Song:2010mg}.  

The anisotropic flow can be described via the Fourier expansion \cite{Poskanzer:1998yz} of the azimuthal angle distribution of the particle production:
\begin{eqnarray}
\label{eq:1-1}
\frac{dN}{d\phi} = \dfrac{N}{2\pi}  \left(  1+2\sum_{n=1}V_{n} e^{-in\phi} \right)  ,
\end{eqnarray}
where $V_{n} = v_{n}\exp(in\Psi_{n})$ is the n$^{\mathrm{th}}$ complex flow vector, $\Psi_{n}$ represents the flow vector direction, and $v_{n}$ is the flow vector magnitude. The azimuthal anisotropic flow harmonic $v_{1}$ is known as directed flow, $v_{2}$ as elliptic flow, and $v_{3}$ as triangular flow, etc.

To a good degree, the lower-order flow harmonics $v_{2}$ and $v_{3}$  are linearly related to the initial-state anisotropies $\varepsilon_{2}$ and $\varepsilon_{3}$ respectively~\cite{Niemi:2012aj}. However, the higher-order flow harmonics, $v_{n>3}$, arising from linear response to the same-order initial-state anisotropies along with non-linear response to the lower-order eccentricities $\varepsilon_{2}$ and/or $\varepsilon_{3}$~\cite{Teaney:2012ke}. Consequently, the full benefit of the higher-order flow harmonics for $\eta/s$ extraction~\cite{Yan:2015jma} benefits form a robust separation of their linear and non-linear contributions.

The higher-order flow harmonic $V_{4}$ can be expressed as:
\begin{eqnarray}\label{eq:1-3}
V_{4}                                     &=&  V_{4}^{ \mathrm{ Linear}} +  V_{4}^{ \mathrm{ Non-linear}}, \\
V_{4}^{ \mathrm{ Non-linear}}  &=&  \chi_{4,22} V_{2}  V_{2},
\end{eqnarray}
where  $\chi_{4,22}$  is the non-linear response coefficients. The value of $\chi_{4,22}$ constrains the magnitude of $V^{ \mathrm{ Non-linear}}_{4}$. Also the magnitude of $V^{ \mathrm{ Non-linear}}_{4}$ encodes the correlations between the flow symmetry planes $\Psi_{2}$ and  $\Psi_{4}$. 

In this work, we employ the multiparticle cumulant method \cite{Jia:2017hbm} to measure the $\pT$-integrated inclusive, non-linear and linear higher-order flow harmonic $v_{4}$ in collisions of U+U at $\sqrt{s_{NN}}$ = 193~GeV, Cu+Au at $\sqrt{s_{NN}}$ = 200~GeV and Au+Au at several beam energies.

\section{Method}\label{Sec:2}

The STAR data of charged particles were analyzed with the multiparticle cumulant technique \cite{Bilandzic:2010jr,Jia:2017hbm}. The framework for the standard cumulant method is discussed in Ref.~\cite{Bilandzic:2010jr}; its extension to the subevents method is reported in Ref.~\cite{Jia:2017hbm}. 
In order to minimize the non-flow correlations in the two-subevent method, the cumulants are constructed from two-subevents which are separated in $\mathrm{\eta}$.  Thus, the constructed multiparticle correlations can be written as:

\begin{eqnarray}\label{eq:2-1}
v_{n} &=&  \langle  \langle \cos (n (\varphi^{A}_{1} -  \varphi^{B}_{2} )) \rangle  \rangle^{1/2},\\ \nonumber
C_{n+m,n,m}           &=&   \langle \langle \cos ( (n+m) \varphi_{1}^{A} - n \varphi_{2}^{B} -  m \varphi_{3}^{B}) \rangle \rangle ,\\ \nonumber
\langle v_{n}^{2} v_{m}^{2}  \rangle &=& \langle \langle \cos ( n \varphi^{A}_{1} + m \varphi^{A}_{2} -  n \varphi^{B}_{3} -  m \varphi^{B}_{4}) \rangle \rangle,
\end{eqnarray}
where, $\langle \langle \, \rangle \rangle$ represents the average over all particles in the event, which are then averaged over event sample, $k, n$ and $m$ are harmonic numbers and $\varphi_{i}$ is the i$^{\mathrm{th}}$ particle's azimuthal angle. 
For the two-subevent method, subevent A and subevent B are required to have a minimum $\Delta\eta > 0.6$ separation, i.e. $\eta_{A}~ > 0.3$ and $\eta_{B}~ < -0.3$.


Using  Eq.(\ref{eq:2-1}) the linear and non-linear modes in the  higher order anisotropic flow harmonic $v_{4}$ can be expressed as:
\begin{eqnarray}\label{eq:2-2}
v_{4}^{ \mathrm{Non-linear}} &=& C_{4,22} / \sqrt{\langle \mathrm{v_2^2 v_2^2 }\rangle}, \\ \nonumber
v_{4}^{ \mathrm{Linear}}      &=& \sqrt{ (v^{ \mathrm{Inclusive}}_{4})^{\,2} - (v^{ \mathrm{Non-linear}}_{4})^{\,2}  }.
\end{eqnarray}

Equation (\ref{eq:2-2}) assumes that the linear and non-linear contributions in $v_{4}$ are independent~\cite{Magdy:2020bhd}, which is a correct approach if the correlation between the lower- ($n=2,3$) and higher-order ($n>3$) flow coefficients  is weak.

\section{Results}
\label{Results}

\begin{figure*}[h!]
  \vskip -0.1cm
 \centering{
\includegraphics[width=0.4\linewidth,angle=-90]{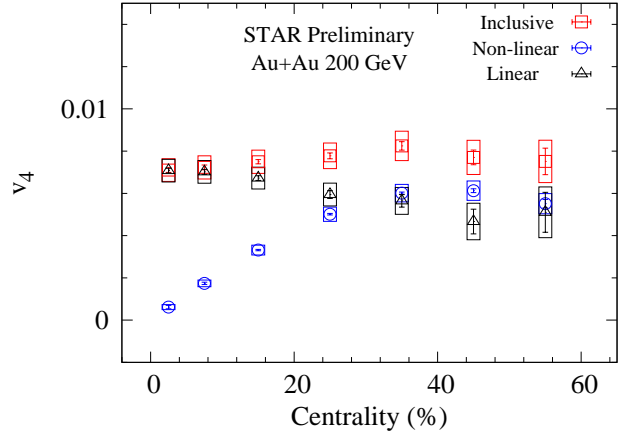}
\vskip -0.2cm
\caption{The inclusive, non-linear and linear higher-order flow harmonic $v_{4}$ using the two-subevent cumulant method as a function of centrality in the $\pT$ range from $0.2$ to $4.0$~\GeVc are shown for Au+Au collisions at $\sqrt{s_{NN}}$=200~GeV. The respective systematic uncertainties are shown as open boxes.
 \label{Fig:1} }}
 \vskip -0.2cm
\end{figure*}

A centrality dependencies of the inclusive, linear and non-linear $v_{4}$  in the $\pT$ range from $0.2$ to $4.0$~\GeVc ~for Au+Au collisions at $\roots$ = 200 GeV are shown in Fig.~\ref{Fig:1}. Our study indicates that the $v^{ \mathrm{Linear}}_{4}$ depends weakly on the collision centrality and it dominates over the non-linear contribution to the inclusive  $v_{4}$ in central collisions.

 \begin{figure*}[h!]
  \vskip -0.2cm
 \centering{
\includegraphics[width=1.0\linewidth,angle=0]{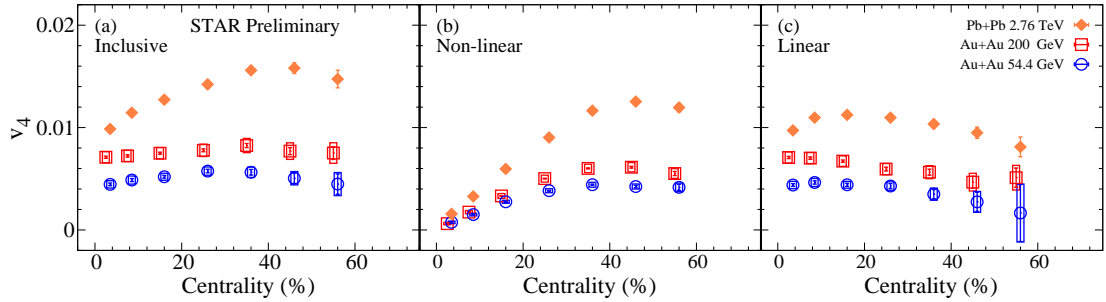}
\vskip -0.2cm
\caption{The inclusive, non-linear and linear higher-order flow harmonic $v_{4}$ using the two-subevent cumulant method as a function of centrality in the $\pT$ range from $0.2$ to $4.0$ \GeVc ~are shown for Au+Au collisions at $\sqrt{s_{NN}}$=200 and 54.4~GeV. The respective systematic uncertainties are shown as open boxes. The results are compared with the LHC  measurements in the $\pT$ range from $0.2$ to $5.0$ \GeVc ~for Pb+Pb collisions at $\roots$ = 2.76~TeV~\cite{Acharya:2017zfg}.
 \label{Fig:2} }}
 \vskip -0.2cm
\end{figure*}

 \begin{figure*}[h!]
  \vskip -0.2cm
 \centering{
\includegraphics[width=1.0\linewidth,angle=0]{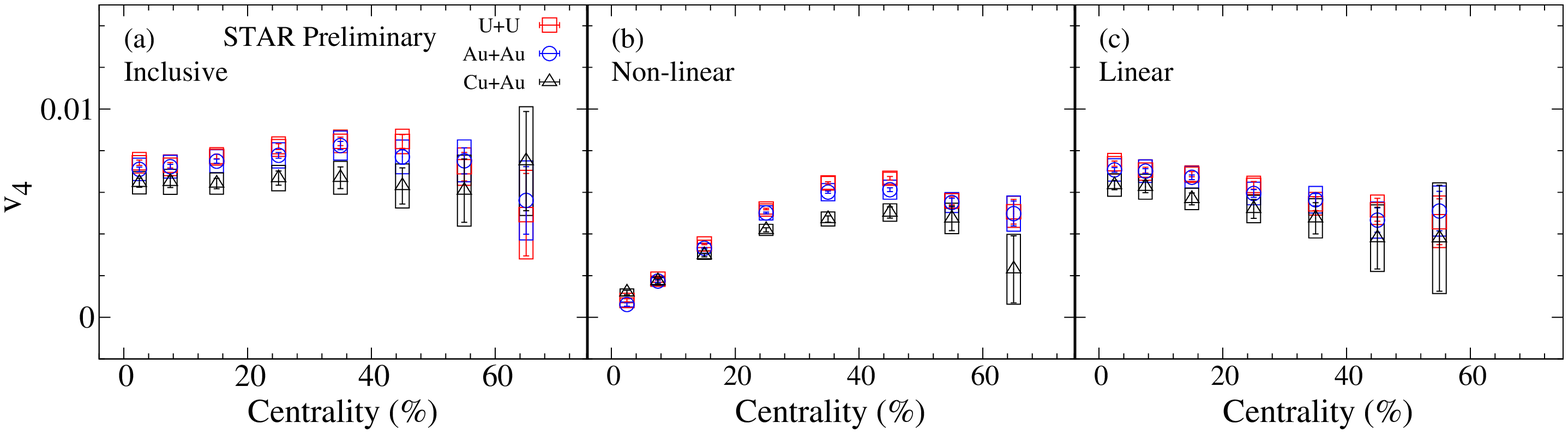}
\vskip -0.2cm
\caption{The inclusive, non-linear and linear higher-order flow harmonic $v_{4}$ are shown for U+U collisions at $\sqrt{s_{NN}}$ = 193~GeV, and Au+Au and Cu+Au collisions at $\sqrt{s_{NN}}$ = 200~GeV. The presented results are measured using the two-subevent cumulant method as a function of centrality in the $\pT$ range from $0.2$ to $4.0$~\GeVc. The respective systematic uncertainties are shown as open boxes. 
 \label{Fig:3} }}
 \vskip -0.2cm
\end{figure*}

Figure~\ref{Fig:2} compares the  centrality dependence of the inclusive, linear and non-linear $v_{4}$  in the $\pT$ range from $0.2$ to $4.0$~\GeVc ~for Au+Au collisions at $\roots$ = 200 and 54.4~GeV. For both energies we observe that the linear mode of $v_{4}$ has a weak centrality dependence, and it is the dominant contribution to the inclusive  $v_{4}$ in central collisions. 
The preliminary results are compared with similar LHC measurements in the $\pT$ range from $0.2$ to $5.0$ \GeVc ~for Pb+Pb collisions at $\roots$ = 2.76~TeV~\cite{Acharya:2017zfg}.
The observed difference of $v_{4}$ magnitudes between Au+Au collisions at $\roots$ =  54.4 and 200~GeV, and Pb+Pb collisions at $\roots$ = 2.76~TeV could be driven by the difference in the viscous effects between those energies.

The preliminary results for U+U collisions at $\sqrt{s_{NN}}$ = 193~GeV, and Au+Au and Cu+Au collisions at $\sqrt{s_{NN}}$ = 200~GeV are shown  in Fig.~\ref{Fig:3}. The magnitudes and trends for both inclusive and non-linear $v_{4}$ show a weak system dependence, albeit with more visible differences between Cu+Au and Au+Au than between U+U and Au+Au.

 \section{Summary}
 \label{summary}
In summary, we have used the cumulant method to measure the inclusive, linear and non-linear $v_{4}$ as a function of collision centrality in  U+U collisions at $\sqrt{s_{NN}}$ = 193~GeV, Cu+Au at $\sqrt{s_{NN}}$ = 200~GeV and Au+Au at several beam energies.  The measurements show the expected characteristic dependence of the inclusive, linear and non-linear $v_{4}$ on centrality, system size and beam energy. Our study indicates that the linear contribution to the inclusive $v_{4}$ dominates over the non-linear contribution  in central collisions for all presented energies and systems. These newly presented measurements may give extra constraints to test different initial-state models and assist to accurate extraction of the QGP specific shear viscosity.




\section*{Acknowledgments}
The author thank Prof. ShinIchi Esumi for the very successful discussions.
This research is supported by the US Department of Energy under contract DE-FG02-94ER40865
%

\section*{References}

\bibliographystyle{elsarticle-num}
\bibliography{ref}

\end{document}